\documentclass[twocolumn,psfig,amssymb]{aa}

\usepackage{graphicx,natbib}
\bibpunct{(}{)}{;}{a}{}{,} % to follow the A&A style
\begin{document}

\title{Hydrodynamic stability and mode coupling in Keplerian flows:
local strato-rotational analysis}
\titlerunning{Hydrodynamic stability and mode coupling in stratified Keplerian flows}

%\author{J.-P. Zahn,\inst{1} A. G. Tevzadze\inst{2} and G. D. Chagelishvili\inst{2}}
\author{A. G. Tevzadze\inst{1}, G. D. Chagelishvili\inst{1} and J.-P. Zahn\inst{2} }
\authorrunning{Tevzadze et al.}

%\offprints{offprints}

%\institute{$^1$ LUTH, Observatoire de Paris, CNRS, Universit\'e Paris-Diderot;
%Place Jules Janssen, 92195 Meudon, France \\ $^2$ E. Kharadze Georgian National
%Astrophysical Observatory, 2a Kazbegi Ave., 0160 Tbilisi, Georgia}

\institute{$^1$ E. Kharadze Georgian National Astrophysical
Observatory, 2a Kazbegi Ave., 0160 Tbilisi, Georgia \\ $^2$ LUTH,
Observatoire de Paris, CNRS, Universit\'e Paris-Diderot; Place Jules
Janssen, 92195 Meudon, France}

\date{Received 31 July 2007; accepted  17 October 2007}

\abstract{} {We present a qualitative analysis of key (but yet unappreciated)
linear phenomena in stratified hydrodynamic Keplerian flows:
\emph{(i)} the occurrence of a vortex mode, as a consequence of
strato-rotational balance, with its transient dynamics; \emph{(ii)}
the generation of spiral-density waves (also called inertia-gravity
or $g\Omega$ waves) by the vortex mode through linear mode coupling
in shear flows.} {Non-modal analysis of linearized Boussinesq
equations were written in the shearing sheet approximation of accretion
disk flows.} {It is shown that the combined action of rotation and
stratification introduces a new degree of freedom, vortex mode
perturbation, which is in turn linearly coupled with the spiral-density
waves. These two modes are jointly able to extract energy from the
background flow, and they govern the disk dynamics in the small-scale
range. The transient behavior of these modes is determined by the
non-normality of the Keplerian shear flow. Tightly leading vortex
mode perturbations undergo substantial transient growth, then,
becoming trailing, inevitably generate trailing spiral-density waves
by linear mode coupling. This course of events -- transient growth
plus coupling -- is particularly pronounced for perturbation
harmonics with comparable azimuthal and vertical scales, and it
renders the energy dynamics similar to the 3D unbounded plane
Couette flow case.} {Our investigation strongly suggests that the
so-called bypass concept of turbulence, which has been recently
developed by the hydrodynamic community for spectrally stable shear
flows, can also be applied to Keplerian disks. This conjecture may
be confirmed by appropriate numerical simulations that take the vertical stratification and consequent mode coupling into account in the high Reynolds number regime.}


\maketitle

\section{Introduction}

According to classical fluid dynamics, unmagnetized disk flows in
Keplerian rotation (more generally: with angular momentum increasing
outward and with no extremum of vorticity) are spectrally stable;
however, there is irrefutable observational evidence that such disks
have to be turbulent.  Due to this apparent contradiction, disk
turbulence is often considered as some sort of mystery.  However, an
analogous dilemma that existed in laboratory/engineering flows has
been solved by the hydrodynamic community in the 90s of the last
century, where a breakthrough was accomplished in the
comprehension of turbulence in spectrally/asymptotically stable
shear flows (e.g. in the plane Couette flow).

Let us briefly recall the essence of that breakthrough  (see
Chagelishvili et al. 2003 for details). Traditional stability theory
followed the approach of Rayleigh (1880) where the instability is
determined by the presence of exponentially growing modes that are
solutions of the linearized dynamic equations. Only recently has one
become aware that operators involved in the the modal analysis of
plane shear flows are not normal, hence that the corresponding
eigenfunctions are non-orthogonal and would strongly interfere
(Reddy et al. 1993). For this reason, the emphasis was shifted in
the 90s from the analysis of long time asymptotic flow stability to
the study of short time behavior. It was established that
\emph{asymptotically/Rayleigh stable flows allow for linear
transient growth of vortex and/or wave mode perturbations} (cf.
Gustavsson 1991; Butler and Farrell 1992; Reddy and Henningson 1993;
Trefethen et al. 1993). This fact incited a number of fluid
dynamicists to examine the possibility of a subcritical transition
to turbulence, with the linear stable flow finding a way to bypass
the usual route to turbulence (via linear classical/exponential
instability). On closer examination,
the perturbations reveal rich and complex behavior
in the early transient phase, which leads to
the expectation that they may become self-sustaining  when there is
nonlinear positive feedback.

Based on the interplay of linear
transient growth and nonlinear positive feedback, a new concept
emerged in the hydrodynamic community for the onset of turbulence in
spectrally stable shear flows and was named \emph{bypass
transition} (cf. Boberg \& Brosa 1988; Butler \& Farrell 1992;
Farrell \& Ioannou 1993; Reddy \& Henningson 1993; Gebhardt \&
Grossmann 1994; Henningson \& Reddy 1994; Baggett et al. 1995;
Grossmann 2000; Reshotko 2001; Chagelishvili et al. 2002; Chapman
2002). The bypass scenario differs fundamentally from the classical
scenario of turbulence. In the classical model, exponentially
growing perturbations permanently supply energy to the turbulence
and they do not need any nonlinear feedback for their
self-sustenance, so the role of nonlinear interaction is just to
reduce the scale of perturbations to that of viscous dissipation. In
the bypass model, nonlinearity plays a key role.  The nonlinear
processes are conservative, but in the case of positive feedback,
they ensure the repopulation of perturbations that are able to
extract energy transiently from the mean flow. The self-sustenance
of turbulence is then the result of a subtle and balanced interplay
of linear transient growth and nonlinear positive feedback.
Consequently, thorough examination of the nonlinear interaction
between perturbations is a problem of primary importance, and the
first step is to search and to describe the linear
perturbation modes that will participate in the nonlinear
interactions.

Such linear transient growth is also at work in rotating
hydrodynamic disk flows; however, the Coriolis force causes a
quantitative reduction of the growth rate  there which delays the
onset of turbulence.  Keplerian flows are therefore expected to
become turbulent for Reynolds numbers a few order of magnitudes
higher than for plane subcritical flows (see: Longaretti 2002;
Tevzadze et al. 2003). The possibility of an alternate route to
turbulence gave new impetus to the research on the dynamics of
astrophysical disks (Lominadze et al. 1988; Richard \& Zahn 1999;
Richard 2001; Ioannou \& Kakouris 2001; Tagger 2001; Longaretti
2002; Chagelishvili et al. 2003; Tevzadze et al. 2003; Klahr \&
Bodenheimer 2003; Yecko 2004; Afshordi et al. 2004 Umurhan \& Regev
2004; Umurhan \& Shaviv 2005; Klahr 2004; Bodo et al. 2005;
Mukhopadhyay et al. 2005; Barraco \& Marcus 2005; Johnson \& Gammie
2005a, 2005b; Umurhan 2006).  By adapting the progress of the
hydrodynamic community to the disks flow,
this research is promising for solving the disks' hydrodynamic turbulence problem.

But it remains to be seen whether this route to turbulence actually
applies to astrophysical disks. Compared to plane shear flows, these
possess two additional properties: differential rotation and
vertical stratification. Separate studies of these factors show that
each exerts a stabilizing effect on the flow: these include
numerical calculation of the stability of unstratified flows by Shen
et al. (2006), experiments on Keplerian rotation without
stratification by Ji et al. (2006), estimates of the growth rates with
stratification by Brandenburg \& Dintrans (2006). However, it appears
that the combined action of differential rotation and stratification
introduces a new degree of freedom that may influence the flow
stability and lead to turbulence at a high enough Reynolds number.
Indeed, it has been shown that strato-rotational flows may exhibit
global instability in bounded domains (Dubrulle et al. 2005).
However, it is probable that local disk dynamics will also lead to
hydrodynamic turbulence. The study of the linear perturbations in
strato-rotational flow in the local limit can be found in Tevzadze
et al. (2003; hereafter T03); it is shown there that the combined
action of rotation and stratification generates an aperiodic vortex
mode, which undergoes nonmodal transient growth. We conjecture that
this transient growth may be the main energy source for the
turbulence in the bypass scenario.

Although the importance of the transient exchange of energy between
perturbations and mean flow has now been realized by most working in
the field, only a few seem aware that another linear process may play
also a crucial role, namely the linear coupling of modes, which
allows for transient exchange of energy between them. As
shown by Chagelishvili et al. (1997a,b) and Gogoberidze et al.
(2004), the energy exchange between modes is inherent to shear flows
(as the transient exchange of energy between the mean flow and
perturbations), and it determines in many respects the diversity of
perturbation modes and, therefore, of nonlinear processes. Once one
fully realizes the role of nonlinear processes in the bypass concept
discussed above, it becomes evident that the neglect of linear mode
coupling may lead to an incorrect picture of nonlinear (and,
consequently, turbulent) phenomena.

There are signs of such linear mode coupling in the simulations
performed by Klahr (2004), Barraco \& Marcus (2005), and Brandenburg
\& Dintrans (2006), but apparently they were not identified as such.
Compressive waves are present in the simulation by Johnson \& Gammie
(2005b), along with vortical perturbations, but their origin is not
recognized, namely linear mode coupling. On this mode coupling,
attention is focused in T03 and Bodo et al. (2005).  The latter
paper studies the linear dynamics of an imposed two-dimensional pure
vortex mode perturbation,  in a compressible Keplerian disk with
constant mean pressure and density. (Two-dimensionality, i.e. the
neglect of cross-disk variation, is only correct  for perturbations
with characteristic scales close to or larger than the
vertical stratification scale of the disk.) Two modes -- a vortex
mode and a spiral-density wave mode -- exist in the system, and they
are strongly coupled. This investigation points out the importance
of mode coupling and the necessity of considering compressibility
for dynamic processes with characteristic scales close to or
larger than the disk thickness.

In T03 we studied the linear dynamics of three-dimensional small-scale perturbations (with characteristic length scales much shorter
than the disk thickness) in compressible, vertically (stably)
stratified Keplerian disks. \emph{The first novelty} presented in
that paper is the occurrence of an interplay between the disk
rotation and stratification. Separately, each of these factors is
stabilizing. However, their interplay gives rise to a new
vortex/aperiodic mode that is able to extract the basic
flow energy  transiently. \emph{The second novelty} is the existence of a linear
coupling of that vortex mode with spiral-density wave (SDW) modes,
that makes SDWs valuable participants of the dynamical processes.
Furthermore, we compared the linear dynamics of the small-scale
perturbations in steady stratified disks with that of the
perturbations in unbounded plane Couette flow. We showed that just
the linear coupling of the disk modes makes the dynamics in the disk
and plane flows similar to each other (provided the Reynolds number
of the disk flow is chosen about three orders of magnitudes higher
than in the plane flow). This similarity suggests that the bypass
concept can also be applied to disk flow and it  further motivates
investigation in this direction.

In this respect, the present paper is a sequel of T03: we
focus again on the mathematical and physical aspects of mode
coupling, while introducing a significant simplification by
neglecting the rotational-acoustic waves. This is justified by the
fact that the characteristic timescale of these modes is much
shorter than for the two other modes,
 when the characteristic lengthscale of the
perturbations is much shorter than the disk thickness. Then the
rotational-acoustic waves are not coupled to the vortex and SDW
modes, and they play a negligible role in the  slow, small-scale
dynamics.  That is why we can cut out the rotational-acoustic waves
(i.e., the flow compressibility) without detriment to the dynamic
picture and confine ourselves to the Boussinesq approximation.
Keeping just vortex and SDW modes, this approximation simplifies
mathematical aspects of the problem and allows an advance in the
analytical description and comprehension of mode coupling.

The paper is organized as follows. In Sect. 2 we introduce the
physical approximations and the mathematical formalism, and describe
the linear strato-rotational balance and the perturbation modes. In
Sect. 3 we present the qualitative and quantitative analysis of the
linear dynamics of perturbations. We summarize and discuss the
results in Sect. 4.

\section{Disk model}

The dynamics of a rotating flow in a central gravity field is
governed by the Navier Stokes equations, which are written here in
the cylindrical coordinates:
\begin{equation}
{\partial \rho \over \partial t} + {\partial \over \partial r}(\rho
V_r) + {1 \over r} {\partial \over \partial \phi} (\rho V_\phi) +
{\partial \over \partial z} (\rho V_z) = 0 ,
\end{equation}
\begin{equation}
{\partial V_r \over \partial t} + ({\bf V} \cdot \nabla) V_r -
{V_\phi^2 \over r}= - {1 \over \rho}{\partial P \over \partial r} -
{\partial \Phi \over \partial r} ,
\end{equation}
\begin{equation}
{\partial V_\phi \over \partial t} + ({\bf V} \cdot \nabla) V_\phi +
{V_r V_\phi \over r} = - {1 \over \rho r}{\partial P \over
\partial \phi} ,
\end{equation}
\begin{equation}
{\partial V_z \over \partial t} + ({\bf V} \cdot \nabla) V_z = - {1
\over \rho} {\partial P \over \partial z} - {\partial \Phi \over
\partial z} ,
\end{equation}
\begin{equation}
\rho = \rho(P,S) ,
\end{equation}
where
\begin{equation}
({\bf V} \cdot \nabla) \equiv V_r {\partial \over \partial r} +
{V_\phi \over r} {\partial \over \partial \phi} + V_z {\partial
\over
\partial z}
\end{equation}
is the component in convective derivative in cylindrical coordinate
system. For the equilibrium state we choose the thin disk approximation
with ${\bf V_0} = (0,V_{0\phi},0)$, with ${\bf \Omega} =
(0,0,\Omega({\bf r}))$ and $V_{0\phi}=r\Omega({\bf r})$. Hence, we
neglect self gravity and assume that the disk is rotationally
supported. Vertical gravity acceleration and stratification
scaleheight can be defined as follows:
\begin{equation}
{\partial \Phi({\bf r}) \over \partial z} = \Omega^2({\bf r}) z
\approx {\rm sgn}(z) g
\end{equation}
\begin{equation}
{1 \over P_0(z)} {\partial P_0(z) \over \partial z} = {1 \over
\rho_0(z)} {\partial \rho_0(z) \over \partial z} \equiv - k_H .
\end{equation}
We consider isothermal equilibrium and study small-scale
perturbations employing  local approximation for further analysis.

\subsection{The shearing sheet formalism in the Boussinesq
approximation}

For the purpose of the linear analysis, we employ local co-rotating
shearing sheet:
\begin{equation}
x \equiv r - r_0, ~~~~~ y \equiv r_0 (\phi - \Omega_0 t), ~~~~~ z =
z ~.
\end{equation}
\begin{equation}
{x \over r_0} , {y \over r_0} , {z \over r_0} \ll 1 ,
\end{equation}
where we neglect the effect of global flow curvature and study the
local influence of the differential rotation on the perturbations
(see Goldreich  \& Linden-Bell 1965). We introduce the linear
perturbations as follows:
\begin{equation}
{\bf V} = {\bf V}_0 + {\bf V}^\prime , ~~~ P = P_0 + P^\prime , ~~~
\rho = \rho_0 + \rho^\prime ~.
\end{equation}
For further simplification we employ the Boussinesq approximation to
neglect compressibility effects. We assume that $k_H={\rm const}$
and restrict analysis to the perturbations with shorter
lengthscales than the vertical stratification scale of the disk
flow. Hence, the system of equations that governs the perturbation
dynamics in the considered limit reduces to
\begin{equation}
\left({\partial \over \partial t} + 2Ax {\partial \over \partial
y}\right) V_x^\prime - 2\Omega_0 V_y^\prime + {1 \over
\rho_0}{\partial P^\prime \over \partial x} = 0 , \label{nsx}
\end{equation}
\begin{equation}
\left({\partial \over \partial t} + 2Ax {\partial \over \partial
y}\right) V_y^\prime + 2(\Omega_0 + A) V_x^\prime + {1 \over \rho_0}
{\partial P^\prime \over \partial y} = 0 , \label{nsy}
\end{equation}
\begin{equation}
\left({\partial \over \partial t} + 2Ax {\partial \over \partial
y}\right) V_z^\prime + {1 \over \rho_0} {\partial P^\prime \over
\partial z} + g {\rho^\prime \over \rho_0} = 0 ,
\label{nsz}
\end{equation}
\begin{equation}
\left({\partial \over \partial t} + 2Ax {\partial \over \partial
y}\right) { \rho^\prime \over \rho_0} - k_H V_z^\prime = 0 ,
\label{nsrho}
\end{equation}
\begin{equation}
{\partial V_x^\prime \over \partial x} + {\partial V_y^\prime \over
\partial y} + {\partial V_z^\prime \over \partial z} = 0 .
\label{nscon}
\end{equation}
Here for brevity $\rho_0 \equiv \rho_0(0)$, and only the $z>0$ case
is considered. We use standard Oort's  constants to quantify the
radial velocity shear:
\begin{equation}
A \equiv {1 \over 2} \left[r {\partial \Omega \over \partial r}
\right]_{r=r_0} \!\!\!\!\!= -{3 \over 4} \Omega_0 , ~~~{\rm and }~~~
A + B = -\Omega_0 .
\end{equation}
Following the standard procedure of nonmodal analysis, we employ
spatial Fourier expansion of perturbations with specific phase
variation:
\begin{equation}
\left\{\begin{array}{c} P^\prime({\bf r},t) \\ \rho^\prime({\bf
r},t)
\\ {\bf V}^\prime({\bf r},t)
\end{array}\right\} =
\left\{\begin{array}{c} p({\bf k},t) \\ \varrho({\bf k},t) \\ {\bf
v}({\bf k},t)
\end{array}\right\}
\exp ( {\rm i}k_x(t)x + {\rm i}k_y y + {\rm i} k_z z),
\end{equation}
where
\begin{equation}
k_x(t) = k_x(0) - 2A k_y t ~.
\end{equation}
This leads to the linear system of ODE that governs the dynamics of
SFH in the considered flow:
\begin{equation}
{{\rm d} \over {\rm d} t}v_x(t) = 2\Omega_0 v_y(t) - {\rm i} k_x(t)
{p(t) \over \rho_0},
\end{equation}
\begin{equation}
{{\rm d} \over {\rm d} t} v_y(t) = 2B v_x(t) - {\rm i} k_y { p(t)
\over \rho_0},
\end{equation}
\begin{equation}
{{\rm d} \over {\rm d} t} v_z(t) = - {\rm i} k_z {p(t) \over \rho_0}
- g {\varrho(t) \over \rho_0} ,
\end{equation}
\begin{equation}
{{\rm d} \over {\rm d} t} {\varrho(t) \over \rho_0} = k_H v_z(t) ,
\end{equation}
\begin{equation}
k_x(t) v_x(t) + k_y v_y + k_z v_z = 0 .
\end{equation}
Introducing new notations for pressure and density
\begin{equation}
P \equiv  {{\rm i} p \over \rho_0} ~,~~~ D \equiv {\varrho \over
\rho_0} ~,
\end{equation}
we obtain an ODE system that does not explicitly contain complex
coefficients. The system is characterized by a temporal invariant
that corresponds to the linear perturbation of the potential
vorticity:
\begin{equation}
W \equiv k_x(t) v_y - k_y v_x + 2B {k_z \over k_H} D = {\rm const.}
\label{invariant}
\end{equation}
With straightforward simplifications
\begin{equation}
P = -g {k_z \over k^2} D + {2 \Omega_0 \over k^2} \left( k_x(t) v_y
- k_y v_x \right) - 4 A {k_y \over k^2} v_x ~,
\end{equation}
we reduce the governing equations to the following ODE system that
is third order in time:
\begin{equation}
{{\rm d} \over {\rm d} t}v_x(t) = ~~~~~~~~~~~~~~~~~~~~~~~~~~~~~~
~~~~~~~~~~~~~~~~~~~~~~~~~~ \label{odex}
\end{equation}
$$ = 2 (\Omega_0+2A) {k_x(t) k_y \over k^2(t)} v_x + 2 \Omega_0
{k_y^2+k_z^2 \over k^2(t)} v_y + g {k_x(t) k_z \over k^2(t)} D ~,
$$
\begin{equation}
{{\rm d} \over {\rm d} t} v_y(t) = ~~~~~~~~~~~~~~~~~~~~~~~~~~~~~~
~~~~~~~~~~~~~~~~~~~~~~~~~~ \label{odey}
\end{equation}
$$
\left( 2B {k_x^2(t)+k_z^2 \over k^2(t)} + 2A {k_y^2 \over k^2(t)}
\right) v_x - 2 \Omega_0 {k_x(t) k_y \over k^2(t)}v_y + g {k_y k_z
\over k^2(t)} D ~,
$$
\begin{equation}
{{\rm d} \over {\rm d} t} D(t) = - {k_H \over k_z} \left( k_x(t) v_x
+ k_y v_y \right) ~, \label{oded}
\end{equation}
with
\begin{equation}
v_z(t) = -{1 \over k_z} \left(k_x(t)v_x(t) + k_y v_y(t) \right) ~.
\label{odez}
\end{equation}
These equations will be employed in the numerical calculations. The
spectral energy of the perturbation SFH can be derived from the sum
of thermal and thermobaric energies (see Eckart 1960), which in the
considered Boussinesq approximation will read as
\begin{equation}
E_k(t) = {\rho_0 \over 2} \left( u_x^2(t) + u_y^2(t) + u_z^2(t) + {g
\over k_H} D^2(t) \right)~.
\end{equation}

\subsection{Rigid rotation limit: strato-rotational balance and
perturbation modes}

In the shearless limit ($A=0$, $B=-\Omega$), the operators that
intervene in the system above become normal, and it is easy to
derive its dispersion relation. Its roots will help us to understand
the mode behavior in the Keplerian flow, although the results are
not applicable as such. Taking the Fourier expansion of the
perturbations in time $\propto \exp({\rm i} \omega t)$, we obtain
\begin{equation}
\omega \left( \omega^2 - 4\Omega_0^2 {k_z^2 \over k^2} - N^2
{k_\perp^2 \over k^2} \right) = 0 ~,
\end{equation}
where
\begin{equation}
N^2 \equiv g k_H = - {g \over \rho_0(z)} {\partial \rho_0(z) \over
\partial z}
\end{equation}
is the Brunt-V\"ais\"al\"a frequency that is constant in our
formalism. The solutions to this dispersion equation consist of two
modes that are generally decoupled in the zero shear limit:

{\it \bf i)} A spiral-density wave (SDW) -- or inertia-gravity wave
-- that describes the oscillations due to the Coriolis force and to
the vertical buoyancy;

{\it \bf ii)} A stationary vortex mode with zero frequency. This
stationary solution describes aperiodic perturbations to the
vertical strato-rotational balance that can be directly derived from
Eqs. (\ref{nsx}-\ref{nscon}):
\begin{equation}
2 \Omega_0 {\partial \over \partial z} {\rm curl} V^\prime = - g
\left( {\partial^2 \over \partial x^2} + {\partial^2 \over \partial
y^2} \right) \rho^\prime ~,
\label{vort1}
\end{equation}
where ${\rm curl} V^\prime$ is the vertical component of the
vorticity:
\begin{equation}
{\rm curl} V^\prime \equiv {\partial V_y^\prime \over \partial x} -
{\partial V_x^\prime \over \partial y} ~.
\end{equation}
We see that stationary vorticity perturbations  in the disk plane
(lefthand side of the Eq. \ref{vort1}) can be sustained by density
perturbations (righthand side of the Eq. \ref{vort1}), provided
that both rotation ($\Omega_0$) and vertical stratification ($g$)
are present. It is the stationary vortex mode, which appears due to
the strato-rotational balance, that becomes dynamically active in
the shear case (for $~A \not=0$) and is then able to extract the
mean flow energy.

\section{Qualitative analysis of the linear dynamics:
transient growth and mode coupling}

The numerical study of SFHs dynamics is governed by Eqs.
(\ref{odex}-\ref{odez}). However, to clarify the physical nature of
the perturbations and their linear dynamics, it is advisable to
rewrite them in another form. Introducing a new variable
\begin{equation}
\Phi(t) \equiv {k(t) \over k_{\perp}(t)} D(t) ~,
\end{equation}
from Eqs. (27),(29)-(31) one obtains the second-order
\emph{inhomogeneous} differential equation
\begin{equation}
\left\{ {{\rm d^2} \over {\rm d} t^2} + \omega_\phi^2(t) \right\}
\Phi(t) = f_\phi(t) W ~,
\label{secondorder}
\end{equation}
with the frequency
\begin{equation}
\omega_\phi^2(t) = N^2{k^2_{\perp} \over k^2} +
4\Omega_0(\Omega_0+A){k^2_z \over k^2}+ ~~~~~~~~~~~~~~~~~~~~~~~~~~~~
\end{equation}
$$
~~~~~~~~~~~~ +\left(8A\Omega_0- 4A^2{k^2_x \over k^2}+12A^2{k^2_y
\over k^2_{\perp}}\right) {k_y^2k_z^2 \over k^2k_{\perp}^2} ~,
$$
and the coupling function
\begin{equation}
f_\phi(t) = - 2 {k_H k_z \over {k k_{\perp}}}\left(\Omega_0 +
2A{k_y^2\over k_{\perp}^2}\right) ~.
\end{equation}

Equation (\ref{secondorder}) is essentially the same as the main
dynamical equation of Chagelishvili et al. (1997), in which one form
of mode linear coupling in shear flows has been described, namely
the generation of wave SFHs by related SFHs of vortex mode
perturbations. Thus one can borrow the qualitative analysis from
that paper. As a result, Eq. (\ref{secondorder}) describes the dynamics of
SFHs of two
different modes of perturbations:\\
\textbf{(a)} the spiral-density wave mode (SDW), which is described
by the general solution of the corresponding homogeneous equation.
Consequently, this wave mode has zero potential vorticity (i.e.
$W=0$, cf. Eq. \ref{invariant}). Note that the frequency
$\omega_\phi(t)$ of the wave SFHs
is time-dependent.\\
\textbf{(b)} the aperiodic vortex mode, that originates from the
equation inhomogeneity $~(f_\phi(t)W)~$ and is associated with the
particular solution of the inhomogeneous equation. The amplitude of
the vortex mode is proportional to $~f_\phi(t)W~$ and tends to
zero with $~W \rightarrow 0$. In other words, the vortex mode has
nonzero potential vorticity and at the same time acquires nonzero
density perturbation.

\noindent Any perturbation at any moment can be decomposed in the sum of a
wave and a vortex mode.

The character of the dynamics depends on which mode SFH is initially imposed
 in Eq. (\ref{secondorder}): pure wave, pure vortex, or a
mixture of the two. Note that the dynamics of tight SFHs
$~(|k_x(0)/k_y| \gg 1)~$ is adiabatic, whereas that of open SFHs
$~(|k_x(0)/k_y| \sim 1)$ is non-adiabatic. Thus a SFH, which
initially is tightly leading $~(k_x(t)/k_y \ll -1)$, becomes open in
due course (with non-adiabatic dynamics), and will finally be
tightly trailing $~(k_x(t)/k_y \gg 1)$.

\begin{figure}[t] %1
\begin{center}
\includegraphics[width=\hsize]{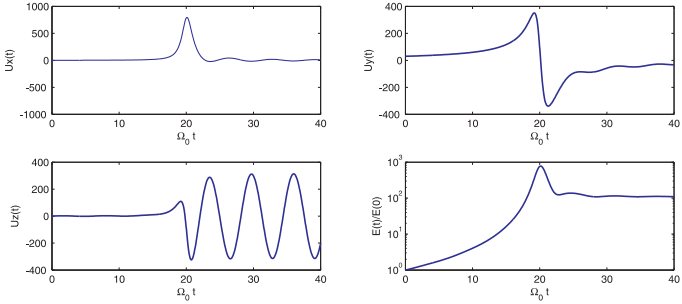}
\end{center}
\caption{The evolution of a single SFH when initial conditions
correspond to the tightly leading pure vortex mode. Velocity
components ($v_x(t)k_H/\Omega_0$, $v_y(t)k_H/\Omega_0$,
$v_z(t)k_H/\Omega_0$,) and normalized energy of perturbation SFH
($E_k(t)/E(0)$) are shown. Here $~k_x(0)/k_y=-30$,
$~k_y=k_z=10k_H$.} \label{Fig_vortex}
\end{figure}

\begin{figure}[t] %2
\begin{center}
\includegraphics[width=\hsize]{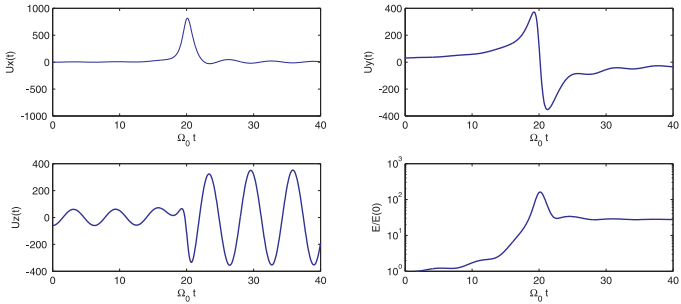}
\end{center}
\caption{Same as Fig. 1, but when the initial conditions for the
single SFH correspond to a mix of vortex and SDW modes. Note that
the differences are rather minor, since the vortex mode
dominates.} \label{Fig_mix}
\end{figure}

To impose initially pure vortex mode perturbations, we use
approximate solutions that can be derived from Eq. (\ref{secondorder}):
\begin{equation}
\Phi_0(t) = {f_\phi(t) \over \omega_\phi^2(t)} W ~.
\label{adiab}
\end{equation}
This solution describes a vortex mode when the temporal variation of
the vortical perturbations is much slower than the SDW oscillations.
Hence, we start our numerical integration of Eqs.
(\ref{odex}-\ref{odez}) at wavenumbers where the above condition is
amply satisfied, i.e., at  $|k_x(0)/k_y| \gg 1$.

Figure \ref{Fig_vortex} displays the evolution of a dynamically active
$(k_y=k_z)$ single SFH when the initial conditions correspond to a
tightly leading pure vortex mode $~(k_x(0)/k_y=-30)$. We see that in
the leading phase only that vortex mode is present. Oscillations
(hence wave SFH) appear in the trailing phase after time
$t=t^*=20/\Omega_0$ when $k_x(t)=0$. The SDW oscillates mainly in
the vertical direction; thus oscillations are the most pronounced in
$v_z$. The graph of the density $D$ is similar to that of $v_z$ and so
is not presented here. The energy graph shows the transient growth
in the leading phase. In the trailing phase, the generated wave SFH
keeps the energy (since SDW do not exchange the energy with the mean
flow). This course of events -- the transient growth and subsequent
coupling -- makes the energy dynamics similar to the 3D unbounded
plane Couette flow case. Note that the tighter the initial leading
SFH  is, i.e. the higher the ratio $|k_x(0)/k_y|$, the stronger
the transient growth. The possible maximum value of $|k_x(0)/k_y|$
is determined by the Reynolds number ($Re$). Consequently, the
amplification factor is determined by $~Re~$ and it can become huge
since in Keplerian disks $Re$ reaches literally astronomical values
$(Re > 10^{10})~$ (see Sect. 3 of T03 for details).

Imposing a leading SFH ($k_x(0)/k_y < 0$) of pure wave mode, we do
not detect any notable transient growth in the wave SFH, or any
generation of the vortex mode: since the potential vorticity is time
invariant (see Eq. \ref{invariant}), the wave SFH, having zero
potential vorticity, is not able to generate the vortex SFH, which
has non-zero potential vorticity. Hence, waves with zero
potential vorticity alone, which were also recently described by
Goodman and Balbus (2001), are not expected to support energetically
hydrodynamic turbulence in Keplerian flows.

Figure \ref{Fig_mix} shows the evolution of another dynamically active
single SFH with $k_y=k_z$, again with $k_x(0)/k_y=-30$, but where
the initial conditions correspond to a mix of vortex and SDW modes.
The initial admixture of the wave SFH does not change the dynamic
picture qualitatively, due to the already mentioned reason, that SDW does
not exchange energy with the mean flow. The energy graph is similar
to Fig. 7 of Brandenburg \& Dintrans (2006), though the linear mode
coupling has not been identified there.

Let us concentrate on the linear dynamics when we initially insert  a
tightly leading SFH of vortex mode perturbation. The dynamics is
then the most complex and the energy exchange the most efficient,
due both to the transient growth of the SFH and to mode coupling
with the SDW. The intensity of these processes depends on values of
$k_x(0)/k_y$ and $k_z/k_y$.

\begin{figure}[t] %3
\begin{center}
\includegraphics[width=\hsize]{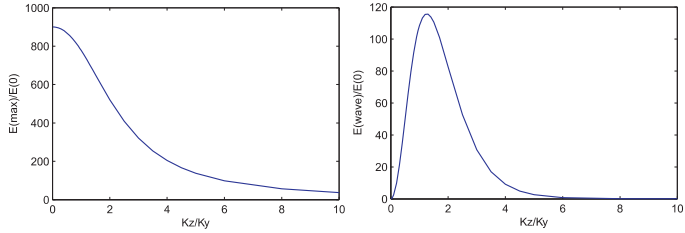}
\end{center}
\caption{{\bf Left}: energy amplification through transient growth
of vortex mode perturbations ${\rm Max}(E(t)/E(0))$ vs. $k_z/k_y$.
~~~~~ {\bf Right}:  mode coupling efficiency measured by the
generated wave energy $E(wave)/E(0)$ vs. $k_z/k_y$. Here
$k_x(0)/k_y=-30$, $k_y=10k_H$ and the maximum efficiency of mode
coupling is achieved at $k_z/k_y=1.25$.} \label{Fig_energy}
\end{figure}

The ratio $E_k(t^*)/E(0)$ gives a good estimate of the nonmodal
growth of vortex SFHs. Here, $t^*$ is the time when $k_x(t^*)=0$.
The spectral behavior (specifically, the dependence on $k_z/k_y$) of
the transient amplification rate calculated by Eqs.
(\ref{odex}-\ref{odez}) is shown in the left panel of Fig.
\ref{Fig_energy}. A constant value of $k_x(0)/k_y=-30$ is chosen to
simulate the energy growth of vortex perturbations at constant
Reynolds number. One can see that the transient amplification is
efficient at low values of $k_z$, while it is strongly reduced for
SFHs with $k_z \gg k_y$.

The right panel of Fig. \ref{Fig_energy} shows the mode coupling
efficiency, measured in units $E_k(wave)/E(0)$, vs $k_z/k_y$, where
$E_k(wave)$ is the final energy of the wave SFH
($E_k(wave)=E(t_{fin})-E(0)$, $t_{fin}=2t^*=40\Omega_0$). That
efficiency has a pronounced maximum at $k_z \simeq k_y$. The energy
of the amplified vortex SFH is transformed into the wave SHF due to
mode coupling, so at later times, the main carrier of the
perturbation energy is the SDW (since vortex decay in the trailing
phase, while waves evolve keeping nearly constant energy).
Therefore, SFHs with $k_z \sim k_y$ are dynamically active and
should play an important role in the turbulent process. These
wavenumbers were chosen in Figs. \ref{Fig_vortex}-\ref{Fig_mix} to
show the dynamical picture of these SFHs and illustrate the
significance of the transient growth and subsequent mode coupling.

\begin{figure}[t] %4
\begin{center}
\includegraphics[width=\hsize]{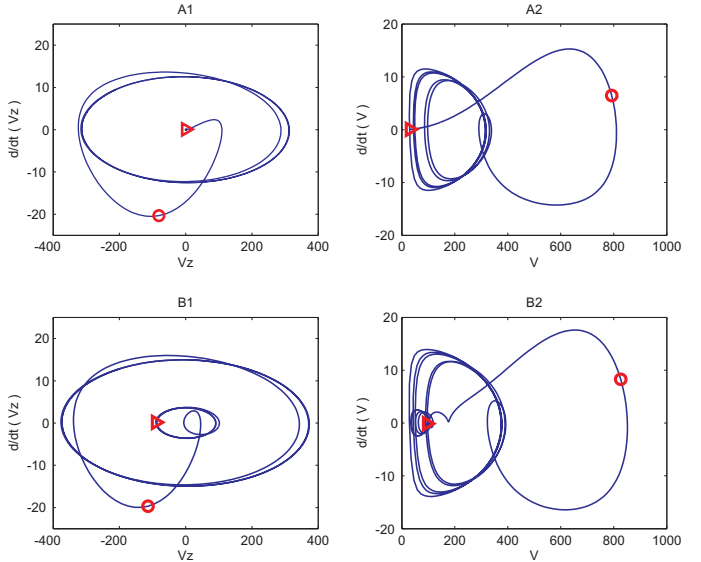}
\end{center}
\caption{The vertical velocity (panels $A.1$ and $B.1$) and total
velocity amplitude (panels $A.2$ and $B.2$) of perturbations in
phase space (${\rm v}$, ${\rm d v /d} t$). Top row shows the
evolution of perturbations when initial conditions correspond to the
vortex mode (panels $A.1$ and $A.2$), while a mix of vortex and wave
modes is shown in the bottom row (panels $B.1$ and $B.2$). A triangle
marks the initial state, while the circle corresponds to the
time when $k_x(t)=0$.} \label{Fig_phase}
\end{figure}

In Fig. \ref{Fig_phase} we have chosen the phase space to display the
evolution of the vertical and total velocities. The initial
conditions correspond to the pure vortex mode (top row, $A.1$ and
$A.2$) and a mix of vortex and wave modes (bottom row, $B.1$ and
$B.2$) presented in Figs. \ref{Fig_energy} and \ref{Fig_phase}. The
initial point is marked by a triangle, and the point at which
$~k_x(t)=0~$ is marked by a circle. The segments from triangle to
circle in the top row graphs describe the transient growth of the
leading phase of the vortex mode. In the trailing phase, after the
circle marker, one can see some subsiding of the vortex harmonic and
generation of the wave (due to mode coupling). The latter is
indicated by the cyclic phase trajectories. The persistence of
cyclic trajectories illustrates the constancy of the wave SFH
energy: the wave is the main carrier of the perturbation energy at
later times. The bottom graphs show that the mix of initial vortex
and wave harmonics evolves much in the same manner. Differences are
seen initially, before the vortex has undergone transient growth.
The graph $B.1$ shows the transient growth and mode coupling in
phase space in terms of the vertical velocity. Note how the initial
``low level'' cyclic phase trajectories are transformed into ``high
level'' cyclic trajectories by transient growth of the vortex
admixture and the subsequent generation of the wave harmonic by the
vortex mode. It illustrates how transient growth raises a
perturbation to a higher energy state.

The vortex mode undergoes the amplification or damping depending on
its phase (leading or trailing) in differentially rotating flows.
From this, it has been speculated (see Johnson \& Gammie 2005) that any
realistic ensemble of these modes covering a wide area in
wave-number space will not exhibit cumulative energy growth.
However, as we have seen in Figs. \ref{Fig_vortex}-\ref{Fig_mix}
total energy of the SFH exhibits asymmetry in growth and decay, due
to the mode coupling. Hence, any ensemble of initial perturbations
that include vortex mode perturbations and cover the region $k_z
\sim k_y$ will undergo overall energy growth in time.

\section{Discussion}

Our motivation in studying mode coupling in the presence of vertical
stratification is quite clear: in unstratified rotating flows the
strato-rotational balance is absent (see Eq. \ref{vort1}),  hence
there will be no vortex mode, whose role is so powerful in
extracting the shear flow energy. Moreover, in stratified rotating
flows mode coupling causes the generation of SDW, which are able to
conserve the extracted energy. Thus one expects astrophysical disks,
which are vertically stratified, to demonstrate intrinsically
different behavior compared to unstratified rotating flows.
Therefore it is not possible to draw conclusions about the stability
of astrophysical disks by considering unstratified rotating flows,
as has been done often in the past and again in recent
investigations: analytical (Balbus 2006), numerical (Shen et al.
2006), and experimental (Ji et al. 2006).

Our analysis shows that, in the local limit, spectrally stable
stratified Keplerian disks allow for two modes of perturbations --
vortex and SDW -- that are jointly able to extract the background
flow energy and determine the disk dynamical activity in the
small-scale range. These modes are linearly coupled due to the
non-normality of Keplerian/shear flow. The coupling is asymmetric:
the vortex mode is able to generate the related  SDW, but the
inverse is not true. This mode coupling is transient (like the energy
exchange between perturbations and the basic flow): the SFH of the
vortex mode generates the wave SFHs during the brief time interval
where it switches from leading to trailing, thus rendering the
dynamics non-adiabatic.

At first sight, the mode coupling described here seems similar to the
phenomenon of geostrophic adjustment, since both lead to the
generation of the spiral density waves. However, these processes are
intrinsically different. Geostrophic adjustment is an initial value
problem that describes the transition from an initially unbalanced
state to that of geostrophic balance. In the general
case, part of the initial conditions containing no potential
vorticity (the wave component) will radiate away as inertial-gravity
waves (see, e.g., Pedlosky 2003) during the geostrophic adjustment,
whereas our initial conditions do not include zero potential vorticity
corresponding to the wave component. Hence, we have
eliminated the wave generation due to the process of the initial
geostropic adjustment.  The waves generated in our case stem from
 linear mode coupling induced by the velocity shear. Moreover,
geostrophic adjustment is mainly a nonlinear process that leads to equilibrium.
 In contrast, wave
generation due to mode coupling is a linear process and does not
describe the relaxation of the system, but hopefully the
opposite: it promotes the transition to turbulence.

The linear dynamics of each leading SFH of vortex mode proceeds in
the following way: initially, the SFH extracts energy from the basic
flow and it grows. At the same time $~k_x(t)/k_y \to -0$. Becoming
trailing, the vortex SFH generates the related SFH of SDW. In what
follows, while tilting (i.e., increasing $~k_x(t)/k_y$), the wave SFH
keeps the energy, whereas the vortex SFH gives its energy back to
the basic flow, so the energy gained  by the leading vortex SFH is
conserved by the SDW. This course of events -- the transient growth
plus coupling -- is strongly pronounced for SFHs with $~k_z/k_y \sim
1$ and makes their energy dynamics similar to that of a 3D unbounded
plane Couette flow.

We are aware that there are differences between them: in the Couette
case, only the vortex mode participates in the dynamical process,
whereas in the disk case, this role is played (in the local limit)
by the symbiosis between vortex and SDW perturbations.
However, given the similarities that we have discussed,  we
conjecture that the bypass concept, which has been developed for the
transition to turbulence in the Couette flow, is also applicable to
rotating stratified disks. One question remains, of course: will the
nonlinear interactions provide positive feedback that is efficient enough
for that bypass transition? To answer it, more numerical simulations
are needed, which must include the vertical stratification and grasp
the mode coupling.

\begin{acknowledgements}

This work is supported by ISTC grant G-1217. A.G.T. and G.D.C. would
like to acknowledge the hospitality of the Observatoire de Paris.

\end{acknowledgements}

%\bibitem[]{}

\end{document}